\documentclass[11pt,fleqn]{article}

\usepackage{amsmath}
\usepackage{amssymb}
\usepackage{authblk}
\usepackage[font=sf]{caption}
\usepackage{float}
\usepackage{graphicx}
\usepackage{hyperref}
\usepackage{lmodern}
\usepackage{makeidx}
\usepackage{siunitx}
\usepackage{wrapfig}

\newcommand*\chem[1]{\ensuremath{\mathrm{#1}}}

\begin{document}
\title{Enhanced photothermal cooling of nanowires}
\author[1]{G.\ Guccione}
\author[2]{M.\ Hosseini}
\author[3]{A.\ Mirzaei}
\author[1]{H.\ J.\ Slatyer}
\author[1]{B.\ C.\ Buchler}
\author[1]{P.\ K.\ Lam\thanks{Ping.Lam@anu.edu.au}}
\affil[1]{Centre for Quantum Computation and Communication Technology, Department of Quantum Science, Research School of Physics and Engineering, The Australian National University, Canberra ACT 2601, Australia}
\affil[2]{Birck Nanotechnology Center, School of Electrical and Computer Engineering, Purdue University, West Lafayette, Indiana 47907, USA}
\affil[3]{Nonlinear Physics Centre, Research School of Physics and Engineering, The Australian National University, Canberra ACT 2601, Australia}

\maketitle

\begin{abstract}
We investigate the optomechanical interaction between light and metallic nanowires through the action of bolometric forces. We show that the response time of the photothermal forces induced on the nanowire is fast and the strength of the interaction can overcome the radiation pressure force. Furthermore, we suggest the photothermal forces can be enhanced by surface plasmon excitation to cool the sub-megahertz vibrational modes of the nanowires close to its quantum limit.
\end{abstract}

\section{Introduction}

% [NEMS and their advantages]
Micro- and nanoscale oscillators have attracted significant attention for sensing applications thanks to their small masses and therefore an extreme sensitivity to the external drives~\cite{Imboden:2014:PhysRep}. To date, nanomechanical devices are used to perform ultra-sensitive measurements of, for example, biological material properties~\cite{Dong:2009:NatNano}, mass sensing at the level of single particles and molecules~\cite{Burg:2007:Nature,Naik:2009:NatNano,Li:2010:NanoLett,Gil-Santos:2010:NatNano}, and sub-attonewton force sensing~\cite{Mamin:2001:ApplPhysLett}. Other fields benefitting from the incredible resolution of nanomechanical sensors include accelerometry~\cite{Liu:1998:JMicroelectromechS}, charge sensing~\cite{Cleland:1998:Nature}, and magnetometry, with tremendous implications for three-dimensional imaging thanks to the resolution at the level of a single electronic spin~\cite{Rugar:2004:Nature}. With interest growing in several areas, and the availability of faster, cheaper, and more precise fabrication techniques, nanoscopic probes are now established for ultra-fast, high-precision sensing in a variety of applications~\cite{Ekinci:2005:RevSciInstrum}.

% [Pushing the sensitivity of NEMS beyond the classical limits]
To push beyond the classical boundaries of sensitivity, the sensors need to operate in the quantum regime. The major challenge is usually represented by the coupling of the system with a thermal environment that hides or even destroys the quantum characteristics of the object. A careful control of the interaction between the sensor and its thermal bath is the key to observing quantum behaviour, especially in the case of mesoscopic objects. Optomechanical systems have been considered recently as viable platforms for revealing the quantum nature of mechanical oscillators at the mesoscale~\cite{Aspelmeyer:2012:PhysToday,Poot:2012:PhysRep,Verhagen:2012:Nature}. Optomechanics allows the coherent and controllable exchange of photonic and phononic excitations and presents a viable option for the neutralization of the incoherent thermal drive of the oscillator. The collective motion of billions of atoms can be regulated by laser light to reveal the quantum properties of the mechanical system. This optomechanics-induced control has been proven to expose the quantum nature of the oscillators by cooling nanomechanical resonators to their quantum ground state (QGS)~\cite{OConnell:2010:Nature,Teufel:2011b:Nature,Chan:2011:Nature}, by generating optomechanical squeezing~\cite{Brooks:2012:Nature,Safavi-Naeini:2013:Nature,Purdy:2013:PhysRevX}, and by achieving quantum state transfer between hybrid systems~\cite{Verhagen:2012:Nature,Palomaki:2013:Nature}.

% [General requirements to reach the QGS]
The first requirement for achieving pure quantum mechanical states of motion is that the mode of vibration be cooled into the QGS by extracting all the energy from the phononic mode and by providing adequate isolation from environmental decoherence. It is, therefore, crucial to be able to control~\cite{Miao:2010:NewJPhys} and characterize the mechanical motion at a level below Heisenberg's uncertainty limit~\cite{Miao:2010:PhysRevA}. The first QGS cooling experiment was performed on a nano-oscillator with a mechanical frequency of $\SI{6}{\giga\hertz}$~\cite{OConnell:2010:Nature}. A general condition for the achievement of the QGS is that the optomechanical system exhibit ultra-high optical and mechanical quality factors. One of the present challenges in the field of cavity optomechanics is to cool the modes of mechanical resonators with smaller frequencies of oscillations to the quantum regime as it entails the fabrication of devices that are not easily accessible with the current technology. For example, the linewidth of the optical resonator might be required on the order of kilohertz or lower. Even if such narrow-band resonators were attained, the interaction bandwidth would be extremely reduced as a result. Very recently, it was proposed that the quantum regime of a specially designed high-stress silicon nitride membrane with sub-megahertz vibration frequency can in principle be reached thanks to the remarkably low mechanical dissipation in such structures~\cite{Norte:2016:PhysRevLett}.

% [Metallic nanowires for plasmon-enhanced interaction]
In this paper, we study an alternative and unconventional approach to the cooling of an oscillator with sub-megahertz vibrational frequencies near the QGS. We suggest the use of metallic nanowires driven by photothermal forces~\cite{Metzger:2004:Nature,Metzger:2008:PhysRevB}, which can be subject to the excitation of plasmonic modes to amplify the interaction between the optical and the mechanical modes. The enhanced photothermal forces can be used to effectively cool nano-sensors without the need for cavity sideband cooling. The tight-light confinement beyond the diffraction limit provided by plasmonic resonances thus enables strong and broadband optomechanical interaction which can be used to probe the quantum limits of photothermal cooling~\cite{De-Liberato:2011:PhysRevA,Restrepo:2011:CRPhys}.

\section{Metallic nanowires as mechanical oscillators}

\begin{figure}[!t]
	\centerline{\includegraphics[width=\textwidth]{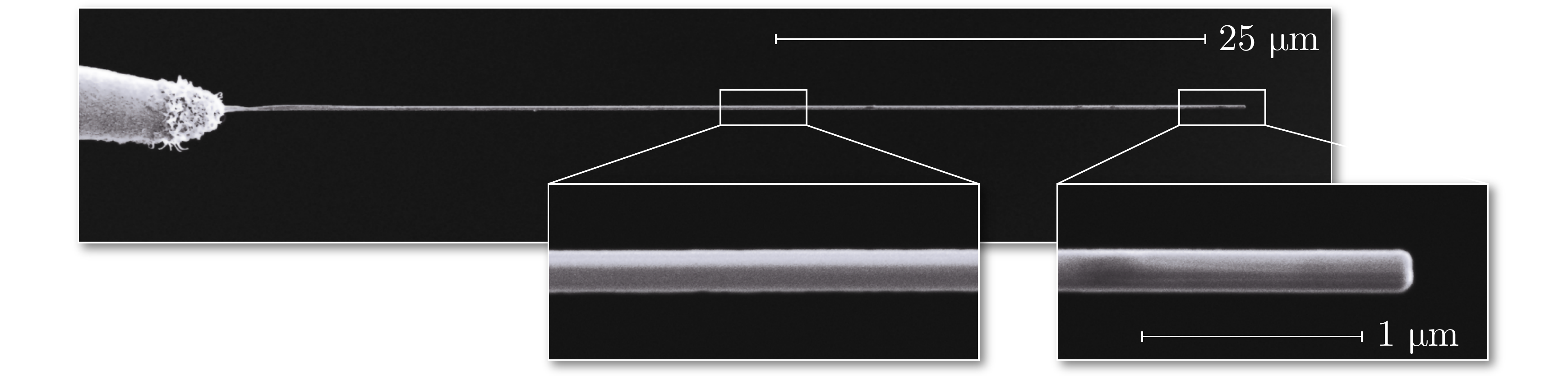}}
	\caption[]{Images of a nanowire, obtained by scanning electron microscopy. The images in the insets show close-ups of the nanowire structure along the shaft and at the tip.}
	\label{fig: nanowires}
\end{figure}

% [Fabrication of the NWs]
The investigations on methods to boost the force sensitivity of nano-probes discussed throughout this paper involve the use of commercial crystalline nanowires\footnote{NN-NCL from NaugaNeedles LLC [\url{http://nauganeedles.com/}]}. Each nanowire is grown coaxially at the extremity of a tungsten needle, by a process that involves dipping of the silver-coated tip of the needle into a droplet of liquid gallium at room temperature. Slow retraction of the tip from the droplet allows the two metals to alloy into a long, uniform rod of \chem{Ag_2Ga} crystallites~\cite{Yazdanpanah:2005:JApplPhys} (cf.\ Fig.~\ref{fig: nanowires}).

% [Applications of the NWs]
Similar nanowires have been used to quantify the surface tension, the viscosity, and other properties of fluids at the microscopic level~\cite{Yazdanpanah:2008:Langmuir}, and to perform high-precision subsurface characterization of nano-structures with high dielectric constants~\cite{Zhao:2010:Nano}. Visual force sensing was also demonstrated by directly monitoring their buckling deformations~\cite{Dobrokhotov:2008:Nano}. In biology, they have been considered for the detection of edge-binding effects in proteins~\cite{Gao:2009:AnalBioanalChem}. In most applications, however, the quality of the measurements in ordinary operating environments is compromised by the thermally induced vibrations of the nanowires. The laser cooling method described here can boost the sensitivity of the nanowires during the transient dynamics of the oscillations~\cite{Hosseini:2014:NatComm}.

% [Specifications of the NWs]
The self-assembled \chem{Ag_2Ga} nanowires have a relatively wide range of specifications, summarized in Table~\ref{tab: nanowire specifications}. They range in size between $\SI{20}{}$ and $\SI{60}{\micro\metre}$ in length and between $\SI{50}{}$ and $\SI{200}{\nano\metre}$ in diameter. To push up the detection efficiency of their vibrational modes, which depends on the light scattered from their surface, a number of specimens were coated with about $\SI{50}{\nano\metre}$ of gold.

\begin{table}[!b]
\begin{center}
\begin{tabular}{ll}
	\multicolumn{1}{c}{\textbf{Quantity}}	&	\multicolumn{1}{c}{\textbf{Value}}							\\\hline\hline
	Length 			&	$\SI{20}{}$--$\SI{60}{\micro\metre}$										\\
	Diameter			&	$\SI{50}{}$--$\SI{200}{\nano\metre}$	 (coated: $\SI{90}{}$--$\SI{500}{\nano\metre}$)	\\
	Density			&	$\SI{8960}{\kilo\gram\per\metre\cubed}$									\\
	Mass			&	$\SI{1}{}$--$\SI{70}{\pico\gram}$ (coated: $\SI{4}{}$--$\SI{150}{\pico\gram}$)		\\
	Oscillation frequency	&	$\SI{20}{}$--$\SI{500}{\kilo\hertz}$ (fundamental)							\\
	Stiffness			&	$\SI{0.1}{}$--$\SI{10}{\milli\newton\per\metre}$ (fundamental)					\\
	Elastic modulus		&	$\approx\SI{100}{\giga\pascal}$											\\
	Damping rate		&	$\SI{0.5}{}$--$\SI{0.9}{\kilo\hertz}$ (in air: $\approx\SI{10}{\kilo\hertz}$)			\\\hline
\end{tabular}
\end{center}
\caption{Typical characteristics of the \chem{Ag_2Ga} nanowires used in the experiments.}
\label{tab: nanowire specifications}
\end{table}

\begin{figure}[!t]
	\centerline{\includegraphics[width=\textwidth]{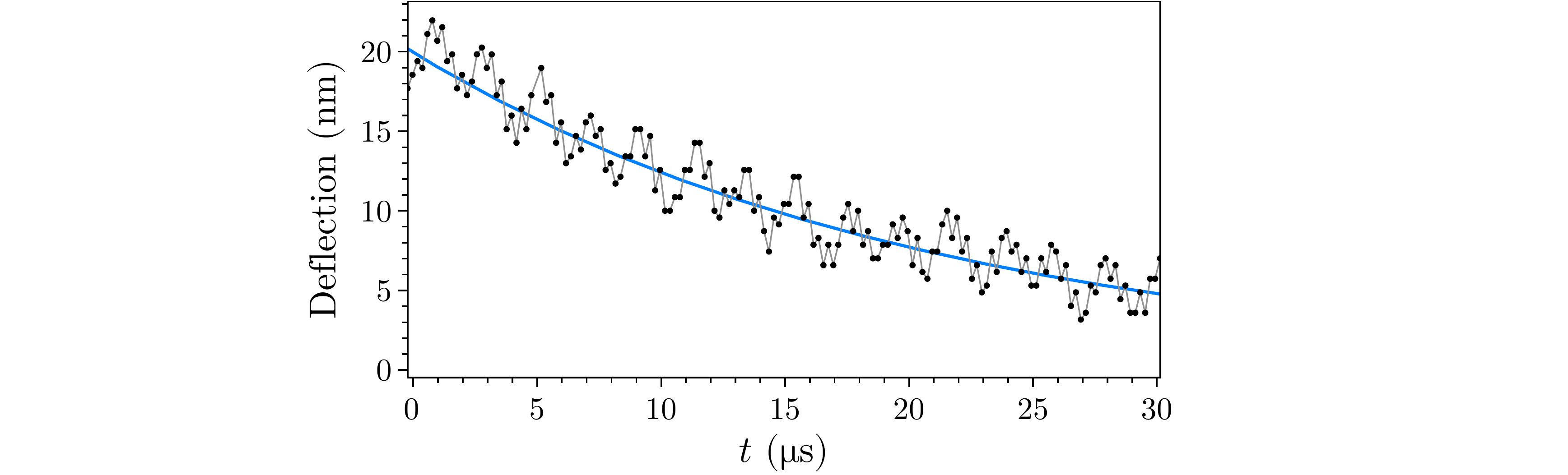}}
	\caption[]{Time-domain measurement of the relaxation rate of a nanowire in air. The nanowire ($\approx\SI{50}{\micro\metre}$ long, $\approx\SI{300}{\nano\metre}$ thick) is subject to a thermally induced deflection until $t=0$, at which point the amplitude of the deflection undergoes exponential decay to the original state. The high-frequency fluctuations on top of the decay represent oscillations at the mechanical frequency. For this specimen the rate obtained by exponential fit of the moving average (solid blue line) is $\SI{7.6(4)}{\kilo\hertz}$.}
	\label{fig: deflection}
\end{figure}

\begin{figure}[!t]
	\centerline{\includegraphics[width=\textwidth]{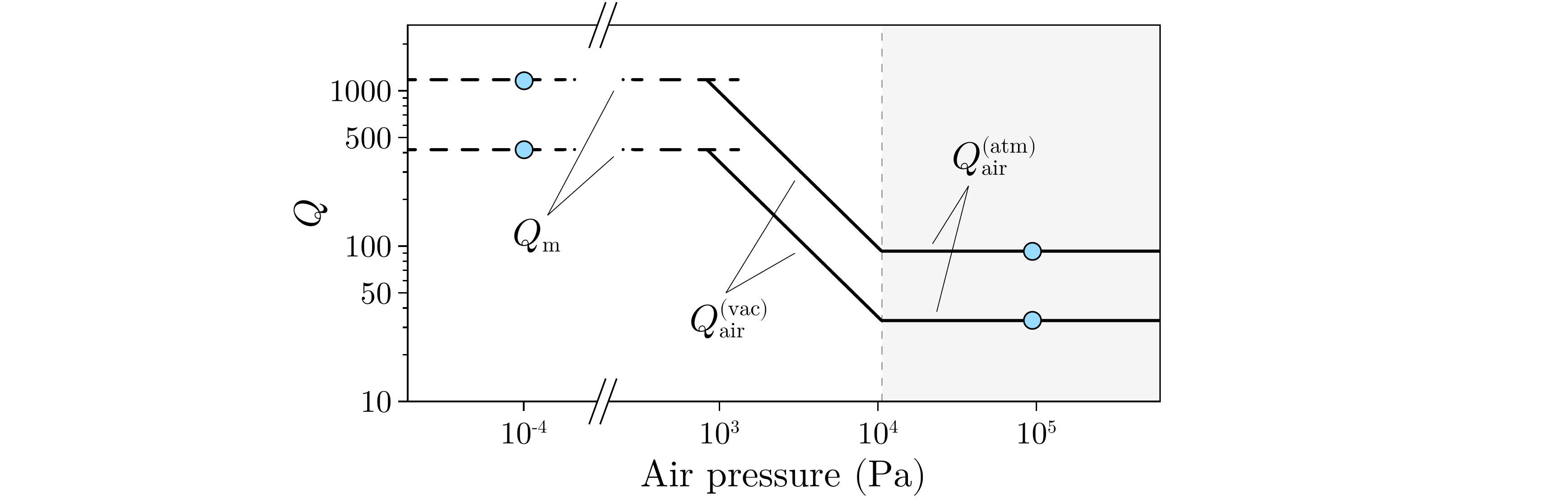}}
	\caption[]{Quality factor of two different oscillation modes in air and in vacuum conditions. The circles correspond to the quality factor obtained by dividing the first two eigenfrequencies of the nanowire by the corresponding damping rates, empirically measured to be $\approx\SI{10}{\kilo\hertz}$ in air and $\approx\SI{0.8}{\kilo\hertz}$ in high vacuum for a nanowire with similar eigenfrequencies. The solid lines indicate the quality factor, dominated by air dissipation, which is expected by the model for a nanowire at room temperature ($T=\SI{300}{\kelvin}$) with elastic modulus $Y=\SI{85}{\giga\pascal}$, density $\rho=\SI{8960}{\kilo\gram\per\metre\cubed}$, diameter $2r=\SI{200}{\nano\metre}$, and length $l=\SI{40}{\micro\metre}$. The molar mass used in the model in the high-pressure regime is $M_\textrm{air}=\SI{28.97e-3}{\kilo\gram\per\mole}$, and the air viscosity required in the high-pressure regime is $\mu_\textrm{air}=\SI{1.8e-5}{\pascal\second}$. The drag coefficient $C_\textrm{d}$, estimated to be between $1$ and $10$ for a cylinder with Reynolds number around unity, was fitted to a value of $2.0$. The transition between the regime of individual gas collisions and viscous dynamics is, for these parameters, around $\SI{10.5}{\kilo\pascal}$. The intrinsic mechanical dissipation takes over at pressures lower than one kilopascal.}
	\label{fig: q factor}
\end{figure}

% [Q factor]
The quality factor $Q$ determines the capability of the system to store energy into the oscillations. It is defined as the ratio of the total energy divided by the energy lost over one cycle. Generally, the quality factor is strongly influenced by a variety of elements, such as thermoelastic and mechanical properties of the oscillator and its support, and the viscosity of the surrounding medium. For a mechanical oscillator where intrinsic mechanical damping and air viscosity are the main factors contributing to the dissipation, the quality factor $Q$ can be expressed as
\begin{align}
	Q = \left(Q_\textrm{m}^{-1}+Q_\textrm{air}^{-1}\right)^{-1},	\label{eqn: quality factor}
\end{align}
where $Q_\textrm{m}=\omega_\textrm{m}/\gamma_\textrm{m}$ and $Q_\textrm{air}=\omega_\textrm{m}/\gamma_\textrm{air}$ are the ratio of the oscillator's eigenfrequency and the damping rate due to the intrinsic mechanical losses or due to the air, respectively. The contribution of air viscosity in ordinary atmospheric conditions is typically dominating for most high-quality resonators~\cite{Chen:2011:JApplPhys}, saturating the quality factor to a value that can be estimated by
\begin{align}
	Q_\textrm{air}^\textrm{(atm)} = \frac{2\alpha^2}{\mu_\textrm{air}C_\textrm{d}l^2}\sqrt{\rho A\,YI},	\label{eqn: Q atmospheric}
\end{align}
where $l$ is the length of the cylinder, $A$ is the cross-sectional area, $\rho$ is the density, $Y$ is the elastic modulus, $I$ is the areal moment, $\alpha$ is a mode-dependent coefficient which is respectively equal to $1.87510$ and $4.69409$ for the first two eigenmodes~\cite{Butt:1995:Nano}, $\mu_\textrm{air}$ is the dynamic viscosity of air, and $C_\textrm{d}$ is the drag coefficient, a function of the Reynolds number and of the oscillator's geometry. By transferring the oscillator into vacuum, the lower density of air molecules is such that they interact with the system without further collisions amongst each other. As more air is pumped out, background gas collisions decrease and the quality factor becomes inversely proportional to the pressure $P$~\cite{Newell:1968:Science}:
\begin{align}
	Q_\textrm{air}^\textrm{(vac)} = \sqrt{\frac{\pi}{2}\frac{R T}{M_\textrm{air}}}\frac{\alpha^2r}{2l^2}\sqrt{\frac{\rho\,YI}{A}}\frac{1}{P},	\label{eqn: Q vacuum}
\end{align}
where $R$ is the universal gas constant, $T$ the temperature, and $M_\textrm{air}$ the molar mass of air. The quality factor cannot be increased arbitrarily, however. At some point further reduction of background gas collisions will have little or no effect, as other intrinsic damping attributes prevail. As these are often specific to the manufacturing process or other details not always easily accessible~\cite{Newell:1968:Science,Imboden:2014:PhysRep}, it is hard to predict what is the highest quality factor achievable by the apparatus without a direct measurement. The decay rates of the thermally excited oscillations of the nanowires in air, measured from the linewidth of the resonances on the spectrum analyser, was observed to be around $\SI{10}{\kilo\hertz}$ due to the interaction with gas molecules (cf.\ Fig.~\ref{fig: deflection}). This corresponds to quality factors of up to $50$ for the fundamental modes. Insertion of the nanowires in a vacuum chamber reduced the damping rates to less than $\SI{1}{\kilo\hertz}$, pushing the quality factors to $500$ or more (cf.\ Fig.~\ref{fig: q factor}). When operating in vacuum, the damping rate of the nanowires was inferred from the time domain evolution of the oscillations to overcome the limit in resolution bandwidth of the spectrum analyser. The chamber was operated in high-vacuum conditions at pressures of $\SI{e-4}{\pascal}$ or lower to avoid air having any role in the damping mechanism.

\section{Scattering and detection}

% [Light confinement for sub-wavelength objects]
Optical measurements play an important role in the detection of small movements of the nanostructures. Light confinement is important to enhance the read-out efficiency. Nanophotonic crystal waveguides~\cite{Chan:2009:OptExp} and cavities~\cite{Favero:2009:OptExp} are widely used in applications that require a tight confinement of the optical mode. Very recently, it was demonstrated that a coupling efficiency for spontaneous emission exceeding \SI{85}{\percent} can be achieved when a single carbon nanotube is coupled to a silicon photonic crystal nanobeam cavity with an ultra-low mode volume~\cite{Miura:2014:NatComm}. Coupling of this strength significantly improves light-matter interactions and enables precise control and read-out, crucial qualities for quantum information and sensing. The light confinement in these systems is however limited by diffraction, which prevents sub-wavelength confinement in the nanophotonic structures. When dealing with objects of $\SI{100}{\nano\metre}$ in radius, effects at the sub-wavelength scale become a significant and inherent part of the system. In this regime radiation pressure is dominated by scattering forces, leading to losses and decreasing the capacity of interaction with the oscillator.

% [Scattering model]
To model the scattering of the metallic nanowires, we follow Mie scattering theory for a sub-wavelength cylinder~\cite{Bohren:2008:WileyVCH}. The solutions are expressed in terms of infinite series of the scattering coefficients $\left\{c_n\right\}_{n\in\mathbb{N}}$, whose value strongly depends on the geometry and refractive index of the object as well as the polarization and the angle of incidence of the field. Notably, any display of absorption is derived from a propagation of the imaginary part of the complex refractive index of the material. We limit our analysis to the case of light normally incident to the axis of the nanowire, with beam width $W$ much larger than the cross-sectional dimensions, i.e.\ $W\gg r$ with $r$ being the radius of the cylinder. The angular distribution of the scattered field is
\begin{align}
	E_\textrm{sca}(\phi) = \sqrt{\frac{2}{\pi\xi}}e^{i\left(\frac{3\pi}{4}+\xi\right)}T(\phi)E_\textrm{in},	\label{eqn: scattered field}
\end{align}
where $\phi$ is the polar angle, $\xi=2\pi r/\lambda$ is a dimensionless ratio between the characteristic length of the object and the wavelength, and $E_\textrm{in}$ is simply the input field. The dependence on the refractive index of the object is implicit in the transfer coefficient $T(\phi)$, which is determined by the scattering coefficients as
\begin{align}
	T(\phi) = c_0+2\sum_{n=1}^{+\infty}c_n\cos(n\left(\pi-\phi\right)).	\label{eqn: scattering transfer coefficient}
\end{align}
The extinction, scattering, and absorption efficiencies, equivalent to the ratio between the effective cross section of each process and the cross-sectional area of the target, are also calculable from the scattering coefficients. They are
\begin{align}
	q_\textrm{ext}	&	= \frac{2}{\xi}\left(\Re(c_0)+2\sum_{n=1}^{+\infty}\Re(c_n)\right),	\label{eqn: extinction efficiency}\\
	q_\textrm{sca}	&	= \frac{2}{\xi}\left(|c_0|^2+2\sum_{n=1}^{+\infty}|c_n|^2\right),	\label{eqn: scattering efficiency}\\
	q_\textrm{abs}	&	= q_\textrm{ext}-q_\textrm{sca}.	\label{eqn: absorption efficiency}
\end{align}
It should be specified that, despite their name, these efficiencies are not bound to unity in Mie scattering theory. As a matter of fact, in many examples the light scattered or absorbed is more than that geometrically incident on the object~\cite{Bohren:2008:WileyVCH}. The efficiencies are needed to infer the amount of radiation pressure force contributing to each process~\cite{Biedermann:2009:Report}. We have that the scattering and absorption components of radiation pressure force are respectively
\begin{align}
	F_\textrm{sca}	&	= \frac{q_\textrm{sca}P_\textrm{in}}{c},	\label{eqn: scattering force}\\
	F_\textrm{abs}	&	= \frac{q_\textrm{abs}P_\textrm{in}}{c},	\label{eqn: absorption force}
\end{align}
in terms of the incident power $P_\textrm{in}$, which is calculated by integrating the intensity of the beam over the irradiated cross-sectional area of the cylinder projected onto the plane perpendicular to the propagation of the field~\cite{Svoboda:1994:OptLett}. The scattering coefficients are the only elements needed to calculate all of these quantities that are still unspecified. The reason lies in the fact that their definition differs depending on whether the polarization of the field is parallel or perpendicular to the cylinder's axis:
\begin{align}
	c_n =
	\begin{cases}
		\dfrac{J_n(\nu\xi)J_n'(\xi)-\nu J_n'(\nu\xi)J_n(\xi)}{J_n(\nu\xi)H_n^{1'}(\xi)-\nu J_n'(\nu\xi)H_n^1(\xi)}	&	\textrm{for parallel polarization,}	\\[3ex]
		\dfrac{\nu J_n(\nu\xi)J_n'(\xi)-J_n'(\nu\xi)J_n(\xi)}{\nu J_n(\nu\xi)H_n^{1'}(\xi)-J_n'(\nu\xi)H_n^1(\xi)}	&	\textrm{for perpendicular polarization.}
	\end{cases}
	\label{eqn: scattering coefficients}
\end{align}
Here, $\nu=n_\textrm{nw}/n_0$ is the ratio between the refractive index of the cylindrical nanowire, $n_\textrm{nw}$, and the one of the surrounding medium, $n_0$. The functions $\left\{J_n\right\}_{n\in\mathbb{N}}$ are the Bessel functions of the first kind and $\left\{H_n^1\right\}_{n\in\mathbb{N}}$ are the Hankel functions of the first kind. The prime indicates differentiation relative to the full argument of the relative function. The derivatives for both classes of functions can be easily computed as the halved difference of the involved functions with preceding and succeeding indices, e.g.\ $J_n'(x)=\left(J_{n-1}(x)-J_{n+1}(x)\right)/2$. For any other polarization the result is obtained by the appropriate linear combination of the different coefficients.

\begin{figure}[!t]
	\centerline{\includegraphics[width=\textwidth]{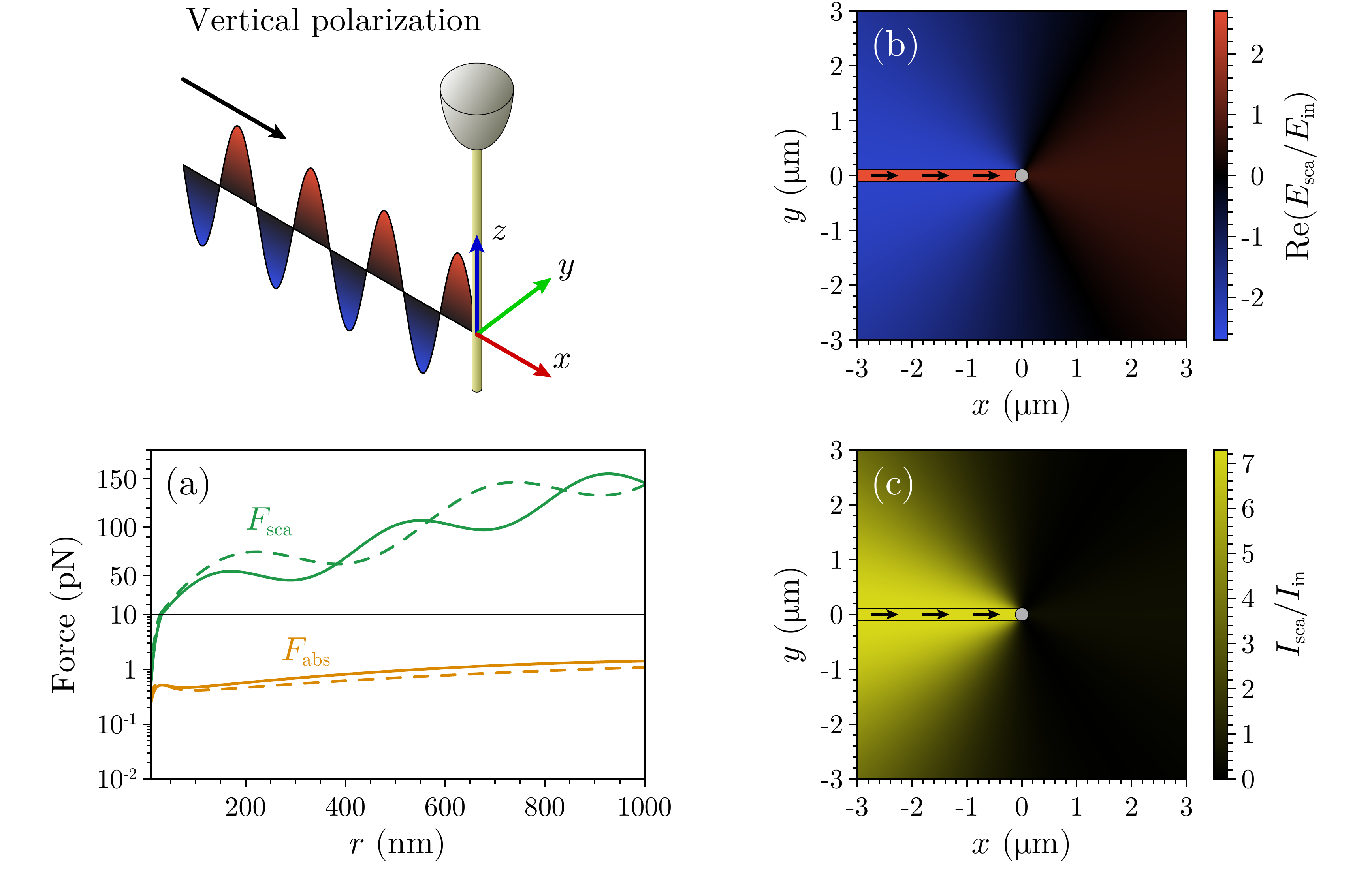}}
	\caption[]{Scattering properties of a nanowire irradiated by a field with polarization parallel to the shaft. The nanowire is assumed to consist entirely of gold, with refractive index $n_\textrm{nw}=0.2+4.8i$ ($0.2+7.0i$) for $\SI{780}{\nano\metre}$ ($\SI{1064}{\nano\metre}$) light. The refractive index of the medium, $n_0$, is taken to be $1$ regardless of whether the nanowire is in air or in vacuum. \;\textbf{(a)} Forces due to scattering (green) and absorption (orange) as a function of the radius of the nanowire, calculated according to Eq.~\ref{eqn: scattering force}--\ref{eqn: absorption force}. The traces are plotted for a beam of width $W=\SI{10}{\micro\metre}$, power of $\SI{50}{\milli\watt}$, and a wavelength of $\SI{780}{\nano\metre}$ (continuous) or $\SI{1064}{\nano\metre}$ (dashed). \;\textbf{(b--c)} Angular distribution of the scattered field and its intensity, as calculated from Eq.~\ref{eqn: scattered field}. The nanowire is placed at the origin and is assumed to have a radius of $\SI{120}{\nano\metre}$. The incident light, of wavelength $\lambda=\SI{780}{\nano\metre}$, is approaching from the negative $x$ axis. Its colour is adjusted to the maximum value of the scale rather than normalized to $1$ in order to reveal the polarization at a glance.}
	\label{fig: scattering (vertical polarization)}
\end{figure}
\begin{figure}[!t]
	\centerline{\includegraphics[width=\textwidth]{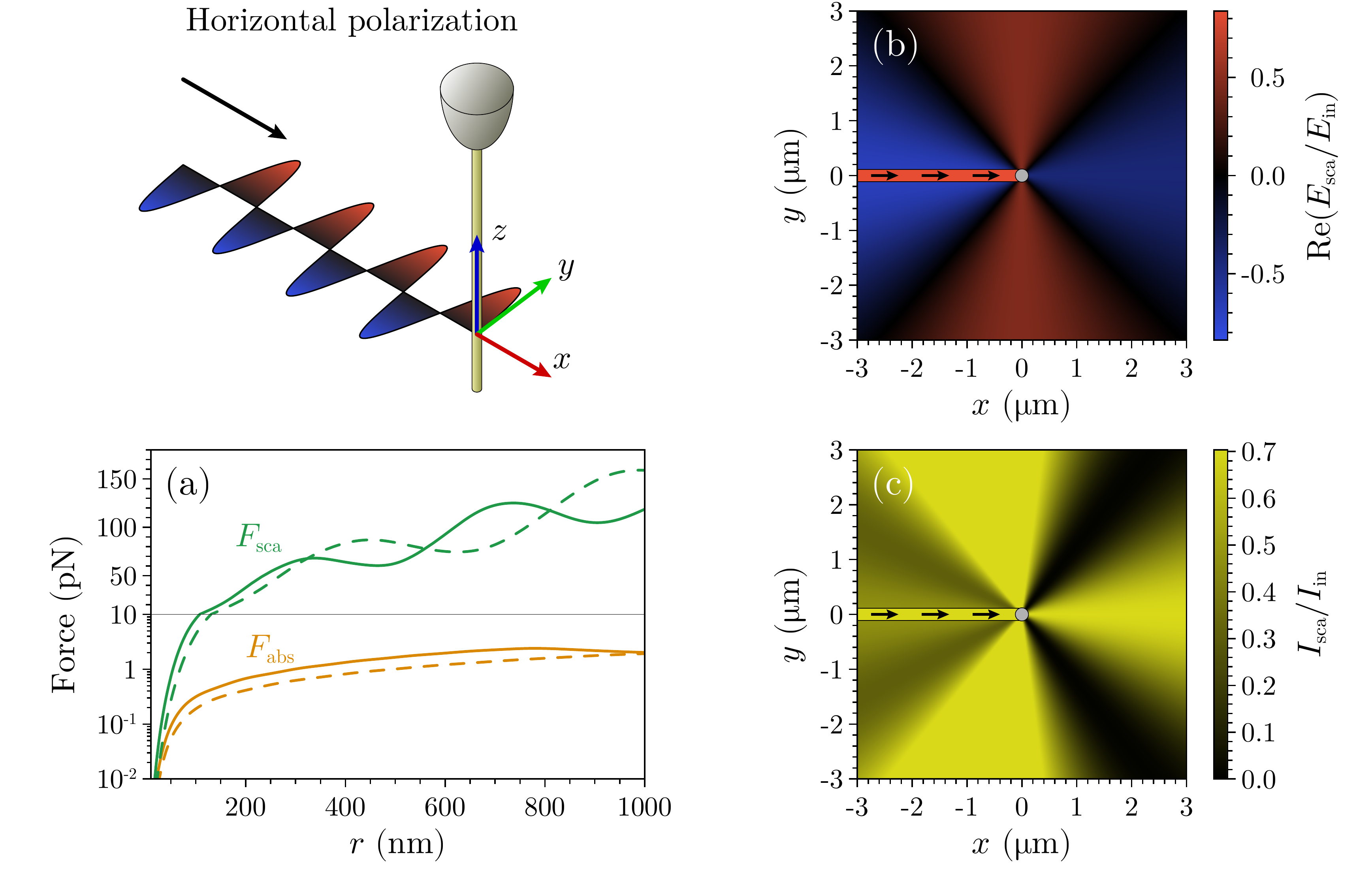}}
	\caption[]{Scattering properties of a nanowire irradiated by a field with polarization normal to the shaft. The parameters used are the same as Fig.~\ref{fig: scattering (vertical polarization)}. \;\textbf{(a)} Forces due to scattering (green) and absorption (orange) as a function of the radius of the nanowire, for a wavelength of $\SI{780}{\nano\metre}$ (continuous) or $\SI{1064}{\nano\metre}$ (dashed). \;\textbf{(b--c)} Angular distribution of the scattered field and its intensity, for incident light of wavelength $\lambda=\SI{780}{\nano\metre}$.}
	\label{fig: scattering (horizontal polarization)}
\end{figure}

% [Scattering results]
The results from this model are presented in Fig.~\ref{fig: scattering (vertical polarization)} for a field with vertical polarization and in Fig.~\ref{fig: scattering (horizontal polarization)} for a field with horizontal polarization. Since the optical properties of \chem{Ag_2Ga} are not very well known~\cite{Biedermann:2009:Report}, all calculations have been performed for a gold-coated nanowire, under the assumption that the effects due to the presence of a different substance at the core could be ignored. These results are only meant for a qualitative analysis aimed at obtaining an order-of-magnitude estimate of the scattering forces and understanding the main directions of the scattered light. The spatial distribution of the scattered field shows a predisposition for backward-scattering, though with quite a wide angle. This is particularly prominent for vertically polarized light, but it holds generally true in other cases as well. It should not be surprising, then, that the most effective procedure for optical detection uses light `reflected' back from the nanowire, although the potential for this (or any other) technique is limited by the aperture and light collecting ability of the setup. Alternatively, one could resort to the `transmission' line instead, looking at the information obtainable from the absence of light in the form of the modulation of the diffracted shadow. Whilst less efficient, this method is not incompatible with the previous one and may be carried out concurrently. As we will see in more detail in the next paragraph, the two detection methods actually address different modes, corresponding to oscillations along orthogonal directions. Looking at the forces acting on the nanowire from Fig.~\ref{fig: scattering (vertical polarization)}a and Fig.~\ref{fig: scattering (horizontal polarization)}a, we see that independently of wavelength or polarization the direct absorption forces are a couple of orders of magnitude smaller than the scattering forces. The model suggests fluctuating values of the forces depending on the radius of the nanowire (an effect less obvious for absorption forces due to the logarithmic scale). The positions and amplitudes of the local minima and maxima depend on the wavelength, creating situations in which the scattering force is, for example, stronger for $\SI{780}{\nano\metre}$ rather than $\SI{1064}{\nano\metre}$. Therefore, depending on the geometry of the nanowire, one wavelength is more suitable for detection while the other is better for external control, as one applies a weaker back action and the other exerts stronger forces. Even though direct absorption forces are relatively small, the indirect effect of photothermal absorption can have much more dramatic consequences than radiation pressure. Optically induced thermal bending of the \chem{Au}/\chem{Ag_2Ga} bimorph nanowires generates a bolometric force which is crucial for the interaction. The bolometric forces observed are estimated to be about one hundred times stronger than the radiation pressure forces estimated by the model.

\begin{figure}[!t]
	\centerline{\includegraphics[width=\textwidth]{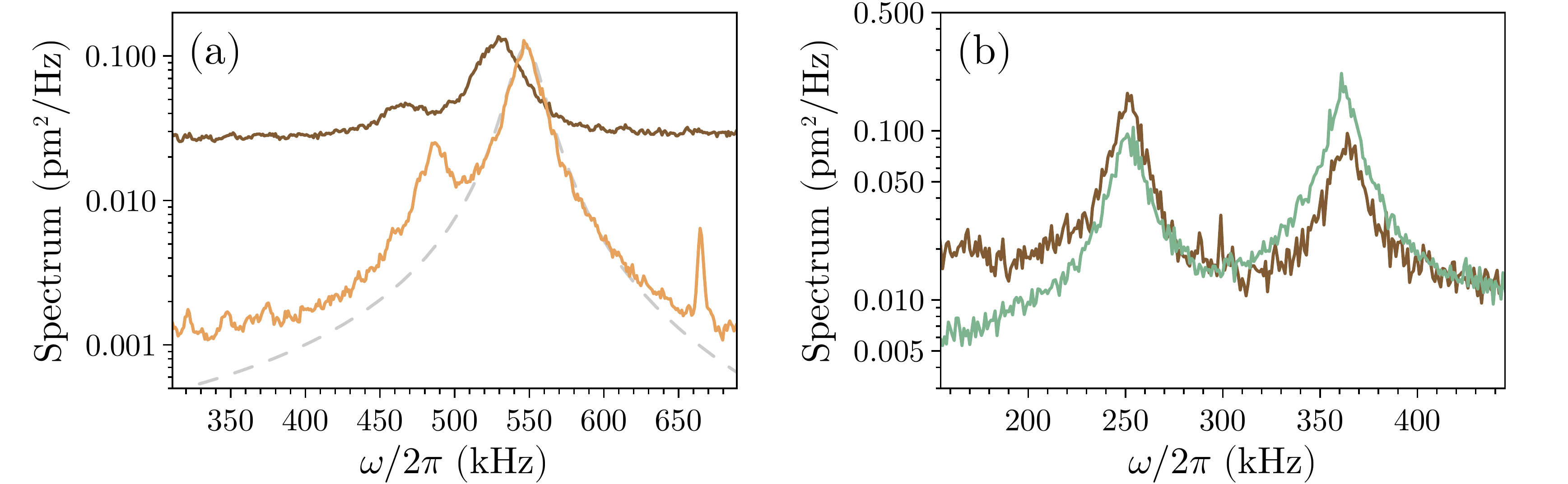}}
	\caption[]{Spectra of the nanowires' thermal fluctuations. All traces have been recorded in ambient conditions. \;\textbf{(a)} Eigenmodes of a nanowire ($\approx\SI{60}{\micro\metre}$ long, $\approx\SI{350}{\nano\metre}$ thick, gold-coated) obtained by interferometry on reflection, once with heterodyne (dark trace) and once with homodyne (light trace). Between the two measurements the nanowire may have been repositioned with a slightly different orientation, accounting for a small variation in the detection ratio of the two modes, and a permanent change was induced by the use of high power, slightly shifting both frequencies. The dashed trace follows the modelled displacement spectrum for an oscillator at room temperature with parameters similar to the measured ones. \;\textbf{(b)} Comparison between the two different detection method: interferometry on reflection (brown) and intensity subtraction from different pixels on transmission (green). The measurements were performed at the same time, and the difference in ratio between the peaks is due to the fact that the two methods have preferential directions of detection. The nanowire is uncoated, $\approx\SI{40}{\micro\metre}$ in length and $\approx\SI{270}{\nano\metre}$ in diameter.}
	\label{fig: detection spectra}
\end{figure}

% [Detection techniques]
After interaction of the light with the nanowire, the light can be measured with various methods to obtain information about the nanowire vibrations. The modes of the nanowire appear in pairs, corresponding to the two motional eigenfrequencies in the plane orthogonal to the cylinder's axis. The slight difference in frequency of each mode originates from imperfections and asymmetries of the nanowire. To detect the vibrations we use two complementary methods applied to the same probe beam. In one method, a split photodiode measures the differences of the diffracted shadow in the transmitted light. Subtracting the photocurrents of the two pixels returns a modulated signal which corresponds to the nanowire fluctuations in the plane orthogonal to the direction of the light propagation. The other method, inspired by Doppler vibrometery~\cite{Biedermann:2010:Nano}, measures the change in phase of the light scattered back by the nanowire by comparing it with a reference beam. This interferometric technique, which collects information on the oscillation along the same direction as the optical axis, proved to be the most practical and efficient. Choosing homodyne over heterodyne detection opens the possibility of phase-locking the interferometer, eliminating any loss of information due to averaging of the beating between the two beams. The two methods are compared in Fig.~\ref{fig: detection spectra}b.

\section{Feedback control}

% [Feedback by bolometric force]
Due to the non-directional Mie scattering from the nanowire, the effective radiation pressure force on the nanowire is small. The dominant optical force on the nanowire in our system is the photothermal bolometric force, estimated to be around one hundred times bigger than the radiation pressure force. The bolometric force is an indirect consequence of optical absorption, arising from the thermal stress and deformation due to the change in temperature. It is particularly consequential for bimorph structures~\cite{Metzger:2008:PhysRevLett,Fu:2011:ApplPhysLett}, where a local increase in temperature induces thermal deflection of different materials at different rates~\cite{Ikuno:2005:ApplPhysLett}.

% [Finite response time of photothermal forces]
While radiation pressure force is almost instantaneous, photothermal forces are typically slow due to the finite thermal conductivity of bulk materials. Nanostructures, however, can exhibit significantly faster response times compared to the bulk structures~\cite{Dhara:2011:PhysRevB, Pradhan:2010:PhD}. Depending on the radius of the nanowire, $r$, and the thermal diffusivity, $\kappa$, the response time $\tau_\textrm{c}\simeq r^2/(4\kappa)$~\cite{Anderson:1983:Science} (required for the central temperature of a local Gaussian temperature distribution to decrease by $\SI{50}{\percent}$) can be as fast as few nanoseconds. This implies a response faster than the mechanical vibrations, which are typically slower than one megahertz. The heat generated by the laser light focused on a small area of the nanowire surface is sufficient to deflect the entire structure~\cite{Ikuno:2005:ApplPhysLett} faster than period of one mechanical oscillation. Therefore, modulating the intensity of the laser light can drive or suppress structural vibrations. The bolometric force may therefore be reliably employed for feedback control purposes. Unintuitively, the fast thermal response time of the nanowires enables the suppression of thermal fluctuations by thermal excitation.

\begin{figure}[!t]
	\centerline{\includegraphics[width=\textwidth]{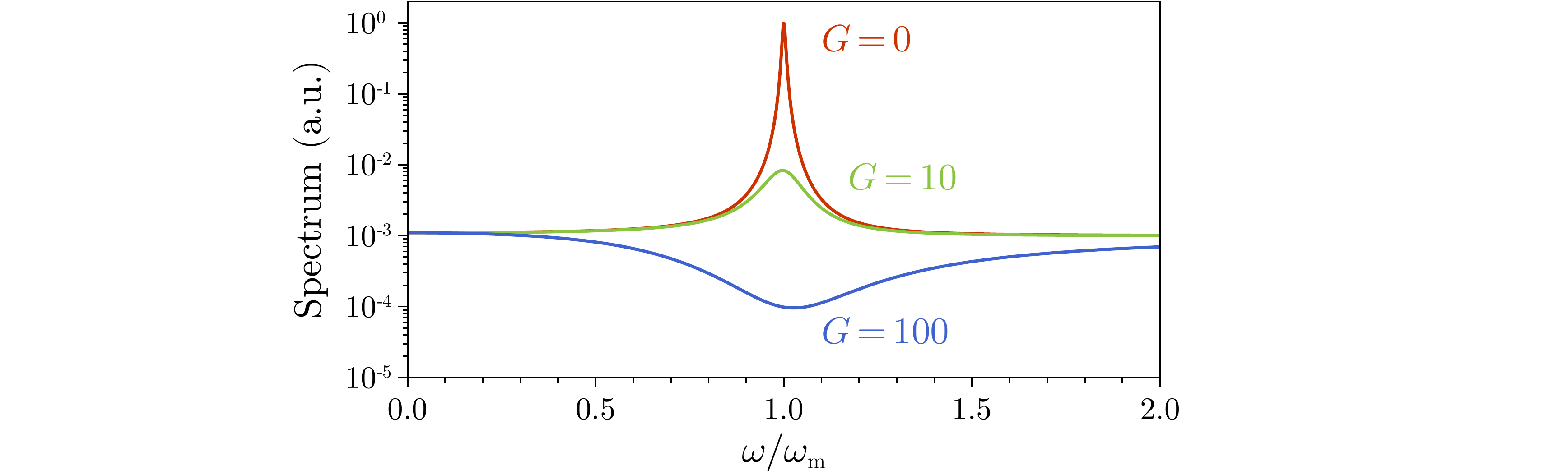}}
		\caption{Theoretical displacement spectra indicative of feedback cooling, for an oscillator with quality factor $Q = 100$. All traces are normalized relative to the value of the spectrum at resonance and with no feedback, and the measurement noise is assumed to be $1000$ times smaller than the thermally driven fluctuations.}
	\label{fig: fb_th}
\end{figure}

% [Theory of negative feedback]
Directing our attention to the case of purely negative feedback, we can study the effect of laser-induced damping and cooling of the system. The effective temperature of a vibrational mode, $T_\textrm{eff}$, is related to the measured spectral density of the fluctuations by the equipartition theorem. Under the effects of active control, in the regime of small feedback gains, the effective temperature is given by
\begin{align}
	T_\textrm{eff} = \frac{T}{1+G},	\label{eqn: low gain effective temperature}
\end{align}
where $T$ is the initial temperature of the bath and $G$ is the feedback gain~\cite{Mancini:1998:PhysRevLett,Pinard:2000:PhysRevA}. The feedback force is equivalent to an apparatus that has an effective damping $\left(1+G\right)\gamma_{m}$, reducing the thermal excitations. Here, $\gamma_\textrm{m}$ is the mechanical dissipation rate of the vibrations in the absence of feedback. For high gain, however, a full treatment of the problem taking into account the detection noise is required. In that regime, it can be shown~\cite{Poggio:2007:PhysRevLett} that the mode temperature is
\begin{align}
	T_\textrm{eff} = \frac{T}{1+G}+\frac{m\omega_\textrm{m}^3}{4k_\textrm{B}Q}\frac{G^2}{1+G}S_\textrm{fb},	\label{eqn: high gain effective temperature}
\end{align}
where $m$, $\omega_\textrm{m}$ are the effective mass and the oscillation frequency of the nanowire, $k_\textrm{B}$ is the Boltzmann constant and $S_\textrm{fb}$ is the noise floor of the displacement measurement, which also modulates the nanowire's motion through the feedback loop. This result shows that the feedback gain cannot be turned up indefinitely to reduce the effective temperature~\cite{Poggio:2007:PhysRevLett}. Therefore, the minimum temperature that can be achieved through feedback cooling by optimizing the gain depends on the measurement noise as
\begin{align}
	T_\textrm{eff}^\textrm{(min)} = \sqrt{\frac{Tm\omega_\textrm{m}^3S_\textrm{fb}}{k_\textrm{B}Q}}.	\label{eqn: minimum effective temperature}
\end{align}
It should be noted that this results from a purely classical argument. To reach the quantum state of the system with active control, it is fundamental for the feedback to operate with quantum-limited detection and to propagate non-classical states~\cite{Wiseman:1993:PhysRevLett}.

\subsection{Experimental results}

\begin{figure}[!t]
	\centerline{\includegraphics[width=\textwidth]{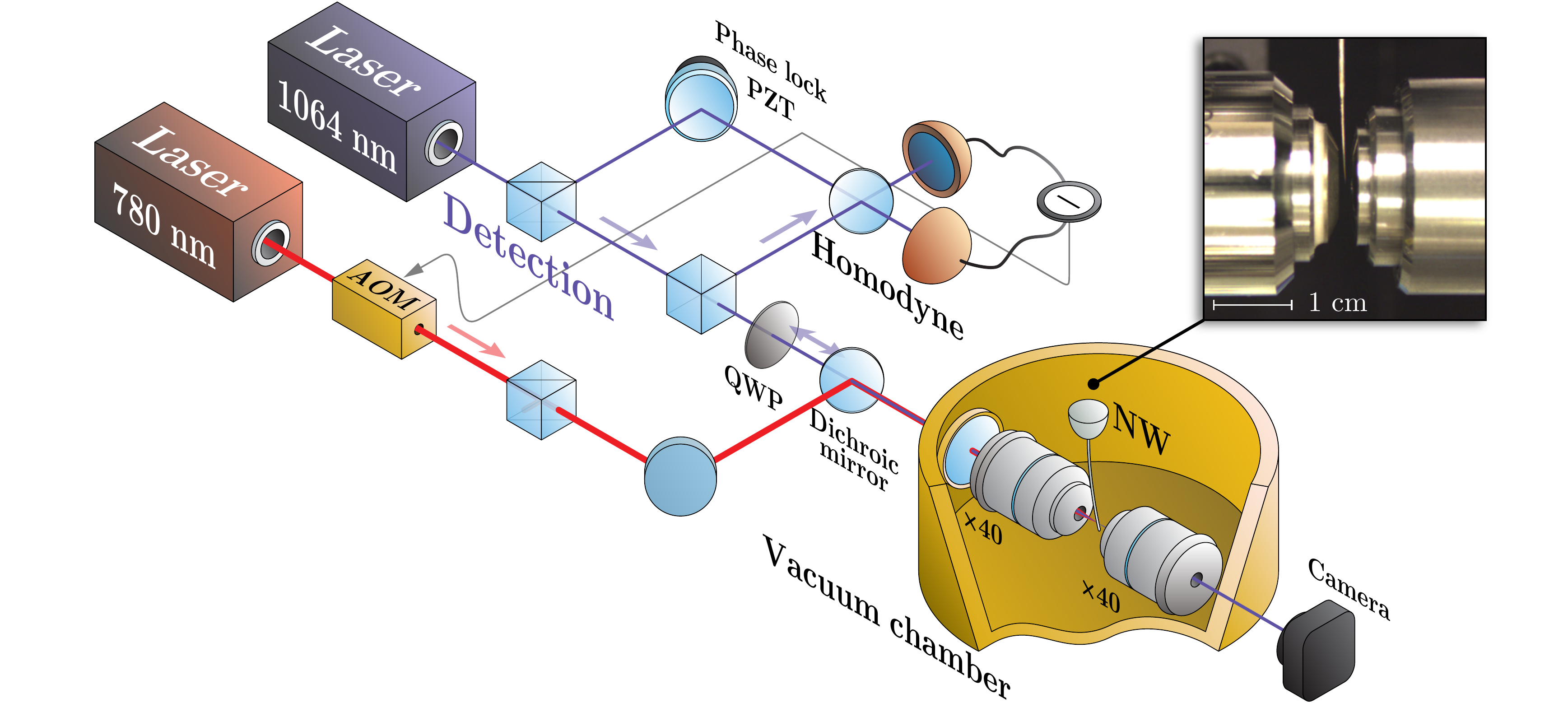}}
	\caption[]{Comprehensive scheme of the experiment on the nanowire (NW). The two optical branches are used for the detection of the thermal motion of the nanowire ($\SI{1064}{\nano\metre}$) and for the feedback control to suppress the thermal fluctuations ($\SI{780}{\nano\metre}$). The inset shows a close-up photograph of the tungsten needle with the nanowire at its tip, between two microscope objective lenses.}
	\label{fig: comprehensive scheme}
\end{figure}

% [Experimental scheme]
In this section we describe results of nanowire detection and its photothermal cooling. Operations on the nanowires have been performed using a two-stage scheme: one part for the detection of the thermally driven modes, one part for the realization of feedback control. The simplified diagram of Fig.~\ref{fig: comprehensive scheme} shows how the two different roles are performed by separate lasers. The requirement for independent sources comes primarily from the necessity of having the beams co-propagate without interference, and the wavelength of each laser is, in itself, only a secondary requirement. The absorption and scattering properties of the nanowires are, of course, elements that need to be considered for an appropriate choice of the operating wavelengths. The detection branch is powered by a $\SI{1064}{\nano\metre}$ beam, while light around $\SI{780}{\nano\metre}$ is used for the actuation, largely because the nanowire under investigation displayed a stronger response in the near infrared and better resistance to high power at longer wavelengths. The vacuum chamber, where the nanowire and the two focusing microscope objective lenses are located, is maintained by an ion pump at pressures of $\SI{e-5}{}$--$\SI{e-4}{\pascal}$, occasionally observed to go down to the $\SI{e-6}{\pascal}$ range. A vacuum-compatible nano-positioning stage is used to allow alignment within the enclosed chamber. In the feedback branch, control on the nanowire's motion is realized by an acousto-optic modulator (AOM) that varies the amplitude of the field and the consequent back-action on the oscillator. The AOM is driven in real time by a signal extracted from the detection scheme, after appropriate processing required to achieve the desired gain and phase for the feedback.

\begin{figure}[!t]
	\centerline{\includegraphics[width=\textwidth]{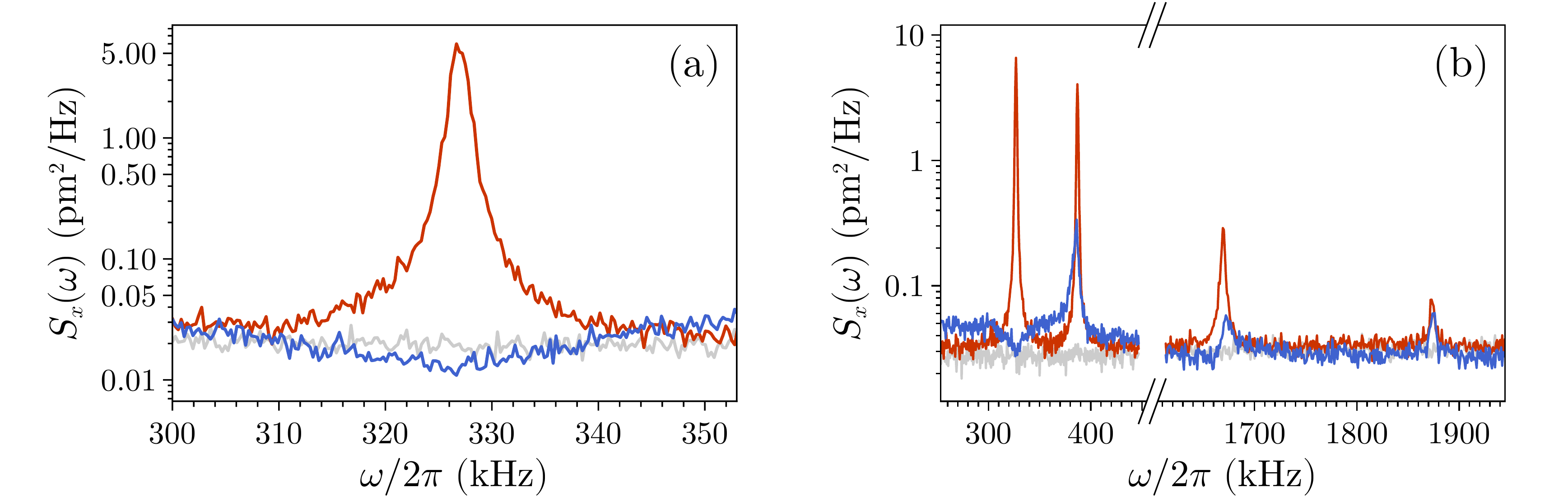}}
	\caption[]{Displacement spectra of a nanowire subject to feedback cooling. The nanowire is $\approx\SI{60}{\micro\metre}$ in length, $\approx\SI{220}{\nano\metre}$ in diameter, and is gold coated. \;\textbf{(a)} Single-mode cooling. The amplitude of the thermal fluctuations (red) is suppressed down to the level of the background noise and beyond (blue), giving rise to squashing in the displacement spectrum. The grey trace indicates the detection noise in the absence of the nanowire. \;\textbf{(b)} Multi-mode cooling. The parameters of the feedback are optimized towards cooling of the mode with the lowest frequency, although in order to obtain cooling of the higher-order modes the phase of the feedback cannot be adjusted optimally and more measurement noise is injected into the system.}
	\label{fig: cooling modes}
\end{figure}

% [Single- and multi-mode cooling]
Examples of the spectral response of the nanowire when subject to feedback cooling are shown in Fig.~\ref{fig: cooling modes}. Feedback control can cool the nanowire's modes both individually and collectively. The practical limits of cold damping imposed by the measurement noise are reached with single-mode cooling, and, for high gain, squashing is observed~\cite{Buchler:1999:OptLett} (cf.\ Fig.~\ref{fig: cooling modes}a). For multi-mode cooling, besides the detection efficiency there are further limits set by the bandwidth of the feedback and more importantly the ability to control its phase across a wide spectrum of frequencies. The technical constraint to the bandwidth scales as the inverse of the characteristic response time $\tau_\textrm{th}$, and is not found to be significant relative to the nanowire's modes. On the other hand, the feedback phase needs to be precisely tuned in order to achieve pure damping. Fine adjustments are only possible over a relatively small frequency range, and pushing more than one mode to the coldest temperature at the same time would only be feasible with the introduction of more advanced controls. Nevertheless, as Fig.~\ref{fig: cooling modes}b shows, the feedback implemented is capable of cooling simultaneously modes spanning up to $\SI{2}{\mega\hertz}$.

\begin{figure}[!t]
	\centerline{\includegraphics[width=\textwidth]{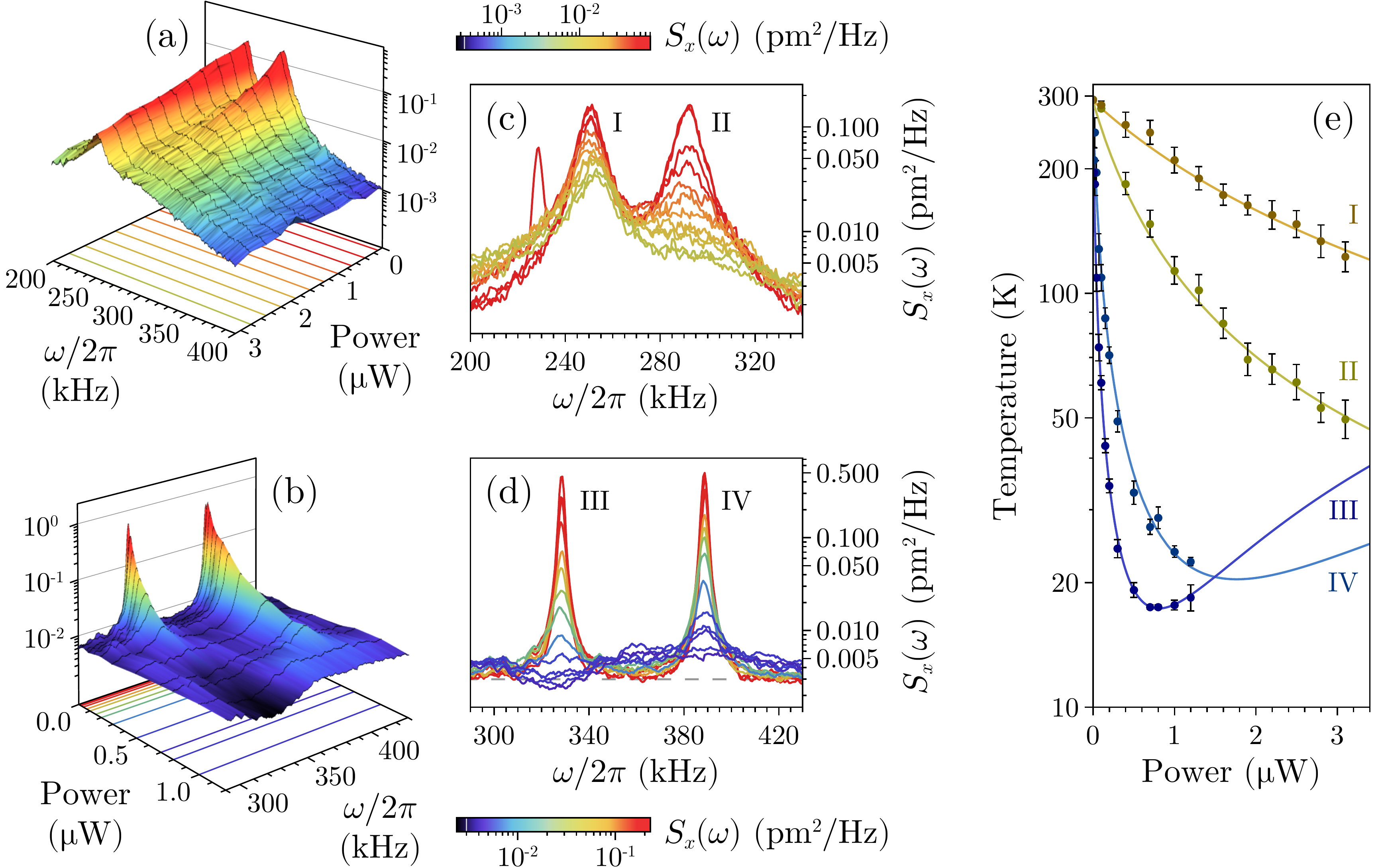}}
	\caption[]{Proof-of-principle experimental result demonstrating thermal noise reduction of a nanowire as a function of gain, which is controlled by the power of the feedback beam. The nanowire used in atmospheric conditions is gold-coated, $\approx\SI{50}{\micro\metre}$ in length and $\approx\SI{300}{\nano\metre}$ in diameter. The nanowire used in vacuum conditions is the same as Fig.~\ref{fig: cooling modes}. \;\textbf{(a--b)} Spectrum of the thermal fluctuations for increasing power, in air (a) and in vacuum (b). The black mesh lines represent the individual traces, projected onto the bottom face with a colour corresponding to the peak value of the coldest mode. The detection noise level is indicated in the colour gradient scale by a grey line. \;\textbf{(c--d)} Front view of the spectra in (a--b), colour-coded according to the peak value of the coldest mode (``II'' in air, ``III'' in vacuum). \;\textbf{(e)} The temperature of each mode, calculated according to Eq.~\ref{eqn: high gain effective temperature}. The error bars are estimated by propagating the uncertainty in the Lorentzian fit of the amplitude noise. The solid lines are theoretical fits assuming a linear relationship between the optical power of the feedback beam and the overall feedback gain. The resulting conversion factors between power and gain estimated for the four modes are $\SI{0.4}{\per\micro\watt}$ (I), $\SI{1.6}{\per\micro\watt}$ (II), $\SI{39.3}{\per\micro\watt}$ (III), and $\SI{15.1}{\per\micro\watt}$ (IV).}
	\label{fig: cooling fit}
\end{figure}

% [Cooling achieved]
Figure~\ref{fig: cooling fit} shows the temperature of the fundamental oscillation modes of the nanowire as a function of the feedback gain applied to cool the thermal vibrations. The results vary considerably depending on whether the nanowire is in ambient or in vacuum conditions, highlighting the importance of a high quality factor for improved results. At atmospheric pressures, the additional dissipation due to the viscosity of the air molecules implies that more power is required to achieve the same levels of actuation obtained in vacuum. At low pressure the quality factor of the oscillations is much higher, rendering the entire procedure more effective. Starting from a room temperature of \SI{293}{\kelvin}, the lowest single-mode temperature attained is $\SI{17.4(2)}{\kelvin}$. In air it was only possible to cool down to $\SI{49(5)}{\kelvin}$. It should be noted that, even under similar pressure conditions, different modes respond to feedback at different rates and one may be cooled more rapidly than the other. The degree of influence is determined by the spatial overlap of the modes with the direction of the bolometric actuation, which does not depend on the orientation relative to the feedback beam. The reduction in temperature and thermal fluctuations achieved is sufficient to increase the measurement sensitivity of the nanowire to impulsive forces~\cite{Hosseini:2014:NatComm}.

% [Possible improvements to the proof-of-principle results]
The results of Fig.~\ref{fig: cooling modes} and~\ref{fig: cooling fit} are primarily limited by the feedback of detection noise into the system and efficiency of the overall control loop. The gain at which the effective temperature corresponds to its minimum value corresponds to the turning point where the measured spectrum shifts from cold damping to squashing~\cite{Buchler:1999:OptLett} (see Fig.~\ref{fig: fb_th}).

\section{Discussion and possible improvements}

% [Limits to the feedback]
A higher quality factor or a thermal bath at lower temperatures, such as in cryogenic conditions, could push the minimum temperature attainable by the feedback to lower limits. Ultimately, however, the results of Fig.~\ref{fig: cooling fit} are limited by the feedback of detection noise into the system and the efficiency of the overall control loop.

% [Improvement by mode confinement]
Cooling of the mechanical resonator to its lowest energy state is only possible with feedback when the measurement sensitivity is increased to be near the standard quantum limit~\cite{Wiseman:1993:PhysRevLett,Clerk:2010:RevModPhys}. This can only be achieved by enhancing the typically weak interaction of the optical field with the mechanical mode of interest. For this reason each photon should be contained within a small volume ($V$), for example by having the nanowire interact with a miniaturized optical resonator. Micro- and nanophotonic structures are capable of reducing the interaction mode volume to values on the order of $\lambda^3$~\cite{Anetsberger:2009:NatPhys,Miura:2014:NatComm}. However, going beyond this regime is not possible with conventional nanophotonics due to the diffraction limit.

% [Improvement by plasmonic interaction]
Alternatively, optomechanical interaction can be enhanced using plasmonic confinement of light. To push the coupling strength beyond what is achievable by conventional nano-optomechanical systems, one can use the metallic properties of the nanowire to take advantage of plasmonic resonances, effectively confining the optical mode below the diffraction limit. To use the plasmon-enhanced absorption to our advantage, the system should be engineered to favour photothermal interaction over scattering forces. In this way, it would be possible to achieve coupling strengths orders of magnitude larger than those achieved via radiation pressure forces~\cite{Thijssen:2013:NanoLett,Thijssen:2015:NanoLett}. 

% [Linking back to the scattering model]
The plasmon effect in our system can be quantified by modelling the scattering and absorption cross sections of the nanowire defined as $\sigma^\textrm{sca}=P_\textrm{sca}/I_\textrm{in}$ and $\sigma^\textrm{abs}=P_\textrm{abs}/I_\textrm{in}$ respectively, where $P_\textrm{sca}$ and $P_\textrm{abs}$ are the scattered and absorbed power and $I_\textrm{in}$ is the intensity of the incident field. These cross sections are related to the scattering and absorption efficiencies (cf.\ Eq.~\ref{eqn: scattering efficiency}--\ref{eqn: absorption efficiency}) and can be obtained by multiplying the corresponding quantity by the cross-sectional area of the nanowire. Therefore, in long nanowires the total value of the cross sections can be represented as a superposition of cylindrical modes. This decomposition can generally be represented as $\sigma^\textrm{sca/abs}=\sum_{n=-\infty}^{+\infty}\sigma^\textrm{sca/abs}_n$, in which $\sigma^\textrm{sca/abs}_n$ is a single-mode scattering or absorption cross section (positive and negative order harmonics) and $n$ is the mode number~\cite{Bohren:2008:WileyVCH}. In a single nanowire, the mode degeneracy of the positive and negative orders appears due to the azimuthal symmetry. The spectral behaviour of every mode is a function of size and material parameters of the nanowire.

% [Corrections required for the plasmonic coupling]
Here we consider the small-size effect in plasmonic materials which may have dielectric constants quite different from those of bulk materials. Confinement of the electrons' motion in conductive nanowires is more significant when the diameter is comparable to the mean free path of the electrons in bulk material, requiring the collision frequency to be modified. This causes a variation of the nanowire's conductivity and optical properties due to the strong influence of the surface. In thinner nanowires, the electrons reach the surface faster and cause an increase in the scattering rate~\cite{Kreibig:1974:JPhysF,Cao:2004:WorldSci}.

\begin{figure}[!t]
	\centerline{\includegraphics[width=\textwidth]{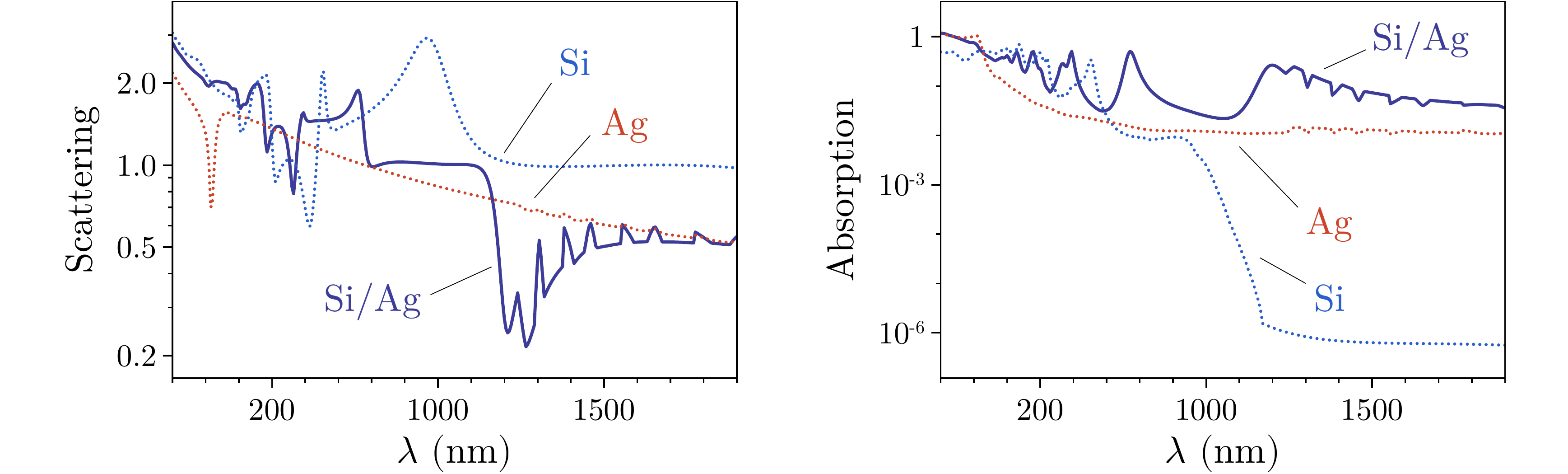}}
	\caption[]{Calculated scattering and absorption cross sections as a function of wavelength of the incident light for three nanowires: silver (\chem{Ag}), silicon ($\chem{Si}$, a non-plasmonic material as reference), and a core/shell \chem{Si}/\chem{Ag} structure. The radius of the silver and the silicon nanowires is, in both cases, $\SI{100}{\nano\metre}$. The plasmonic shell of silver, with a thickness of $\SI{20}{\nano\metre}$, is added to the $100$-\SI{}{\nano\metre} silicon nanowire to reveal the effects of the plasmonic coating on a dielectric nanowire. The cross sections, calculated in two-dimensional simulations, are expressed in unit length of the nanowires and are normalized relative to $2\lambda/\pi$.}
	\label{fig: spr}
\end{figure}

% [Drude's model]
To take this size-dependent effect into account, we calculate the correction to the dielectric permittivity using Drude's model,
\begin{align}
	\epsilon = \epsilon_\infty-\frac{\omega_\textrm{p}^2}{\omega^2+i\gamma\omega},
\end{align}
where $\omega=2\pi c/\lambda$ is the optical angular frequency, $\omega_\textrm{p}$ and $\gamma$ are the plasma angular frequency and the electron collision frequency~\cite{Ordal:1985:ApplOpt}, and $\epsilon_\infty$ is the relative permittivity in bulk material at high frequencies, in which case electrons cannot follow the excitation. To consider the small-size effect, we replace the collision frequency $\gamma$ with one that accounts for the small dimensions of the object, $\gamma_\textrm{nw}=\gamma_\textrm{bulk}+Av_\textrm{F}/d$, where $A=1$, $v_\textrm{F}$ is the Fermi velocity, and $d$ is the characteristic size of the metallic structure (in this case, the thickness of the plasmonic cell)~\cite{Kreibig:1974:JPhysF,Cao:2004:WorldSci}. To obtain the corrected dielectric permittivity of metals in the nanostructure using experimental data, first we use the real and imaginary parts of the bulk material's permittivity to achieve exact $\gamma_\textrm{bulk}$ and $\omega_\textrm{p}$, using Drude's formula as follows:
\begin{align}
\gamma_\textrm{bulk}=\frac{\omega\Im{(\epsilon)}}{(\epsilon_\infty-\Re{(\epsilon)})},	&&	\omega_\textrm{p}^{2}=\omega^{2}\frac{(\epsilon_\infty-\Re{(\epsilon)})^{2}+\Im{(\epsilon)}^{2}}{\epsilon_\infty-\Re{(\epsilon)}}.
\end{align}
Then, by calculating $\gamma_\textrm{nw}$ and using it in Drude's model, we recover the value of $\epsilon$ (one also can use the approximation $\epsilon_\textrm{nw}=\epsilon_\textrm{bulk}+i[\omega^2_\textrm{p}v_\textrm{F}]/[\omega^3d]$). For longer wavelengths these discrepancies can be reduced by using appropriate values of $\epsilon_\infty$, $\gamma_\textrm{bulk}$ and $\omega_\textrm{p}$. Calculated absorption and scattering cross sections are plotted in Fig.~\ref{fig: spr} for \chem{Ag} and \chem{Si} nanowires as a function of the wavelength of the incident beam. The parameters used in Drude's model for silver are: $\epsilon_\infty=1$, $\omega_\textrm{p}=2\pi\times\SI{2.18}{\peta\hertz}$, and $\gamma=2\pi\times\SI{4.353}{\tera\hertz}$.

% [Simulation results]
The simulation results show that metallic nanowires or general metal coated optomechanical structures can be tuned close to their plasmonic modes by appropriate choice of laser wavelength to obtain a hybrid plasmon-optomechanical system. In particular, the detection can be improved by choosing a wavelength favouring scattering over absorption, while the actuation can be enhanced by choosing a wavelength at which most of the energy is absorbed rather than dispersed with the scattered field. As shown in Fig.~\ref{fig: spr}, this complementary response can even be engineered on non-plasmonic systems, such as silicon nanowires, by introducing an appropriate metal coating. Increased absorption can also be realized by tailoring a multi-layered structure on the nanowire~\cite{Mirzaei:2015:Nanoscale}. These novel systems offer a new regime of light-matter interaction, capable of reaching optomechanical interaction strengths much larger than what is available with the current systems~\cite{Thijssen:2013:NanoLett}. Also, the broadband plasmonic interaction is not related to the enhancement obtained when coupling the oscillator to a nanocavity, and it can be used to introduce additional Purcell enhancement~\cite{Exter:1996:PhysRevA}.

% [Typical difficulties of plasmonic systems]
There are, however, typically two main difficulties in using conventional plasmonic systems for quantum experiments: firstly, the plasmonic materials introduce enhanced incoherent absorption and therefore losses; secondly, the metallic structures containing highly confined plasmonic modes suffer from weak coupling to the far field. The loss in plasmonic structures is almost unavoidable, and it has been theoretically shown that the dispersive coupling for the lateral motion of the plasmonic structure is negligible compared to the dissipative coupling~\cite{Hassani-nia:2015:PhysRevA}. Unconventionally, we propose using the dissipative coupling which results in localized heating to our advantage, by engineering a system where photothermal forces are the dominant type of optomechanical interaction, as described above. On the other hand, the weak coupling to the far field can be resolved by interfacing the metallic nanowires with nanophotonic structures~\cite{Miura:2014:NatComm}. Proper design of the structure may lead to optomechanical coupling strength as high as $\SI{2}{\tera\hertz\per\nano\metre}$~\cite{Thijssen:2013:NanoLett}.

\section{Towards quantum ground state cooling}

% [Outlook]
Indulging in speculation, we regard the possibility of reaching the quantum regime of the modes of a mechanical oscillator by deliberate heating. A realistic outlook would comprise enhanced interaction and detection capabilities at the quantum limit. We envision a system where the tip of a \chem{Si} nanowire is placed inside the evanescent mode of a nanophotonic cavity confining the probe beam (of wavelength $\lambda_\textrm{p}\approx\SI{1000}{\nano\metre}$) to amplify the signal-to-noise ratio of the measurement. In this scheme, the control beam (of wavelength $\lambda_\textrm{c}\approx\SI{300}{\nano\metre})$ impinges on a small region of the nanowire away from the tip where the nanowire is coated with silver. The small interaction volume provided by the nanophotonic structure can reduce the noise floor of the displacement measurement to sub-\SI{}{\femto\meter\hertz\tothe{-1/2}} as previously demonstrated~\cite{Schliesser:2009:NatPhys}. This is close to the quantum zero-point fluctuations of the oscillator represented by $x_\textrm{ZPF}=\sqrt{\hbar/\left(2m\omega_\textrm{m}\right)}$, where $\hbar$ is the Planck constant. By additionally choosing the probe and control wavelengths to respectively coincide with plasmonic resonances for scattering and absorption, the nanowire position can be measured to its ultimate precision limit and its thermal motion can be suppressed with high accuracy to allow, in principle, cooling of the vibrational modes of a nanomechanical oscillator near the QGS.

% [Estimates for best cooling]
The temperature required to obtain a mean phonon occupation number $n_\textrm{m}=(e^{\frac{\hbar\omega_\textrm{m}}{k_\textrm{B}T}}-1)^{-1}$ less than one, and therefore reach the QGS of the oscillator, is estimated to be about $\SI{10}{\micro\kelvin}$ for a vibrational frequency of $\SI{200}{\kilo\hertz}$. Taking into account the strong feedback force and a displacement measurement at the quantum limit, an oscillator with a mass of $\SI{10}{\pico\gram}$ and a moderate quality factor of $\SI{e4}{}$ is capable of reaching the vibrational QGS. For example, starting from a cryogenic temperature of $\SI{4}{\kelvin}$, a phonon occupation number of $10$ can be obtained with a displacement noise spectral density of about $\SI{0.1}{\femto\meter\hertz\tothe{-1/2}}$, and less than $1$ phonon is obtained at $\SI{0.01}{\femto\meter\hertz\tothe{-1/2}}$. Nevertheless, one may demand a higher quality factor not only to facilitate the process but also to improve the coherence of the system. The number of coherent oscillations before the assimilation of a phonon from the thermal environment is dictated by the product of the quality factor and the frequency of the oscillation as $Q\cdot\omega_\textrm{m}\times\hbar/\left(k_\textrm{B}T\right)$~\cite{Nguyen:2013:ApplPhysLett,Norte:2016:PhysRevLett}. With a cryogenic bath and the sub-megahertz frequency considered, a quality factor greater than $\SI{4e5}{}$ is required to maintain the coherence in the system for more than one oscillation. In this regime, quantum sensing of impulsive external forces is enabled even at low frequencies.

\section*{Acknowledgements}
This research is supported by the Australian Research Council Centre of Excellences CE110001027, the Discovery Project DP150101035. PKL is supported by the ARC Laureate Fellowship FL150100019, BCB by the ARC Future Fellowship FT100100048.

%\bibliographystyle{custombiblio}
%\bibliography{refs}

\end{document}